\documentclass[
    ,final            % use final for the camera ready runs
  ]
  {aipproc}

\layoutstyle{8x11double}

\begin{document}

\title{The~Flow~Of~Granular~Matter Under~Reduced-Gravity~Conditions}

%%<Replace this text with PACS numbers; choose from this list:
%%                \texttt{http://www.aip..org/pacs/index.html}>}

\author{P.~G.~Hofmeister}{
  address={TU Braunschweig, Institut f\"{u}r Geophysik und extraterrestrische Physik,
			Mendelssohnstr. 3, 38106 Braunschweig, Germany}
}

\author{J.~Blum}{
  address={TU Braunschweig, Institut f\"{u}r Geophysik und extraterrestrische Physik,
			Mendelssohnstr. 3, 38106 Braunschweig, Germany}
}

\author{D.~Hei\ss{}elmann}{
  address={TU Braunschweig, Institut f\"{u}r Geophysik und extraterrestrische Physik,
			Mendelssohnstr. 3, 38106 Braunschweig, Germany}
}

%\author{<author3>}{
%  address={<common address for author2 and author3>}
%  ,altaddress={<author1 address>} % additional visiting address
%}

\begin{abstract}
	To gain a better understanding of the surfaces of planets
 and small bodies in the solar system, the flow behavior of granular
 material for various gravity levels is of utmost interest. We performed a set of 
 reduced-gravity measurements to analyze the flow behavior of granular matter with a
 quasi-2D hourglass under coarse-vacuum conditions 
 and with a tilting avalanche box. We used the Bremen drop tower and a small centrifuge 
 to achieve residual-gravity levels between $0.01~g_0$ and $0.3~g_0$. Both experiments were carried out with
 basalt and glass grains as well as with two kinds of ordinary sand.
For the hourglass experiments, the volume flow through the orifice, the repose and friction angles, and the flow 
behavior of the particles close to the surface were determined. In the avalanche-box experiment, we measured the duration 
of the avalanche, the maximum slope angle as well as the width of the avalanche as a function of the gravity level.

% This template file shows how to use the \texttt{aipproc} class to
% produce a paper with the correct layout for \emph{%
%   AIP Conference Proceedings  8.5in x 11in double column}.
% A full description of the features supported by the \texttt{aipproc}
% class can be found in the \texttt{aipguide.pdf} document accompanying
% the distribution.
% Frequently asked questions can be found in the \texttt{FAQ.txt}
% document.
\end{abstract}

\maketitle

%%%%%%%%%%%%%%%%%%%%%%%%%%%%%%%%%%%%%%%%%%%%
%% MAINMATTER
%%%%%%%%%%%%%%%%%%%%%%%%%%%%%%%%%%%%%%%%%%%%

\section{Background}
All small planetary bodies about which remote-sensing or in-situ surface analyses are available show the occurrence 
of a layer of regolith. This regolith is a loose granular matter derived from meteoritic impacts and the subsequent production 
and re-accumulation of fine granular material on the surface. Due to the lower gravity levels on the surfaces of Mars-, Moon-, or asteroid-size 
bodies, the flow behavior of the regolith under a variety of environmental conditions is fairly unknown. We therefore performed 
granular-matter experiments under reduced-gravity levels to get some insight into the $g$-dependence of surface- and volume flows of granular assemblages.

\section{Particle Properties and Experimental Technique}
As analogs to planetary regolith, which mainly consists of silicates, we chose
samples of basalt spheres, glass spheres, and irregular sand. Table \ref{tab1}
lists the main properties of these samples.
%To investigate the flow behavior of the particles in Table \ref{tab1},
%we developed two setups, which will be explained in more detail below. 
The first experiment releases, upon onset of the low-gravity phase, a small avalanche and allows the measurements of the duration, 
the maximum
\begin{table}[htbp]
\caption{Overview of the particles used in the low-gravity flow experiments.}
 \begin{tabular}{rrr}
%   \hline
   \tablehead{1}{r}{b}{Sample}
  & \tablehead{1}{r}{b}{Size (range) [mm]}
  & \tablehead{1}{r}{b}{Shape}   \\
%   \hline
   Basalt & 0.4 - 0.6 & spherical \\
   Glass & 0.1 - 0.2 & spherical \\
   Sand I & 0.2 - 0.6 & irregular \\
   Sand II & 0.1 - 0.3 & irregular \\
%   \hline
 \end{tabular}
\label{tab1}
\end{table}
slope angle and the width of the resulting avalanche. The second experiment allows to observe the flow of granular matter through a small orifice in a quasi-2D hourglass setup. Here, the slope angles of the outflow and inflow parts, the volume flow, and the surface-flow properties of individual particles can be studied.

\section{The Formation of Reduced-Gravity Conditions}

Artificial zero gravity of 4.7~s was created by dropping the experiments in the drop tower Bremen, a 110~m evacuated tube, inside a pressurized 
experiment container. To reach the desired residual acceleration for the experiments of 0.3, 0.1, 0.03, and 0.01~$g_0$ ($g_0 = 9.814 \rm ms^{-2}$), the experimental setups were mounted on a centrifuge with a diameter of 0.6~m and rotation axis parallel to the earth's gravity. The centrifuge was operated at a well-defined speed so that a constant centrifugal acceleration could be obtained during the 4.7 s free-fall.

\section{\label{exp1}Avalanche-Box Experiments}

\begin{figure}
  \includegraphics[width=0.95\columnwidth]{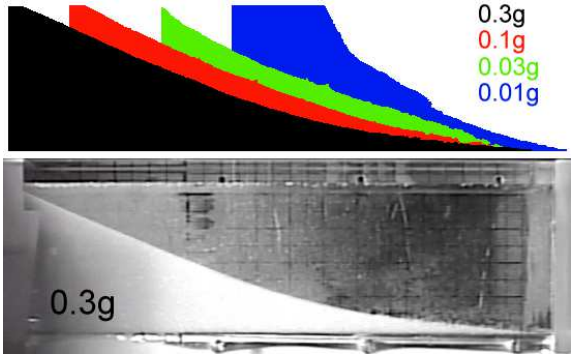}
  \caption{Avalanches performed with glass grains at different gravity levels.
  			The top figure shows binarized images from avalanges at different $g$-levels at the end of the drop-tower
			flight (for clarity shifted in the horizontal direction), the bottom image shows the an 
			original view as recorded
			by the video camera.}
  \label{glasbild}
\end{figure}

The avalanche box has an inner dimension of $4\times4\times15~\rm cm^{3}$.
 Three walls are made of aluminum and three of acrylic glass. The box
 was mounted upright on the centrifuge so that, due to the gravity of the earth prior to the drop, 
 the granular matter was located at the bottom of the box. At the moment
 the experiment is released to free fall, the avalanche is launched by the
 vanishing gravity of the earth and the dominance of the remaining centrifugal force.
The avalanche was recorded with a black-and-white PAL video camera, whose images were subsequently binarized to ease evaluation. In the following, we 
will restrict ourselves to the analysis of the experiments carried out with glass beads.
The top image of Fig. \ref{glasbild} shows the binarized
final images of the respective avalanches at gravity levels of $0,01$, $0.03$, $0.1$, and $0.3~g_0$, whereas the bottom image shows an original snapshot 
of the final avalanche at $0.3~g_0$.

%\begin{figure}
%  \includegraphics[width=1.0\columnwidth]{glasbild}
%  \caption{Avalanches performed with glass grains at different gravity levels. The first four
%			images show binarized images at the end of the drop-tower flight, the bottom image shows the original view as recorded
%			by the video camera.}
%  \label{glasbild}
%\end{figure}

\begin{figure}
  \includegraphics[height=0.95\columnwidth,angle=270]{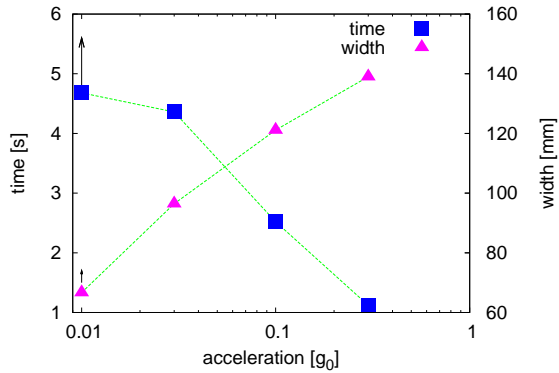}
  \caption{Width of the expired avalanche and time from launch until
		expiration as a function of gravity for glass beads. At the lowest gravity setting
		(0.01~$g_0$), the avalanche did not expire within the flight duration of 4.7~s.
		Therefore, the data points denote the values at end of flight and arrows indicate
		estimated values beyond 4.7~s.}
  \label{timewidthglas}
\end{figure}

Analysis of the image sequences shows that the time from launch to expiration of the avalanche
 increases with decreasing gravity, whereas the width of the expired avalanche increases with increasing gravity level  (see Fig. \ref{timewidthglas}).

%	The resulting slope of the surface of the expired (steady)
% avalanche is shown in figure \ref{glasbild}.
%	The dependence of the general properties on gravitation
% is shown in figure \ref{timewidthglas}

\begin{figure}
  \includegraphics[height=0.95\columnwidth,angle=270]{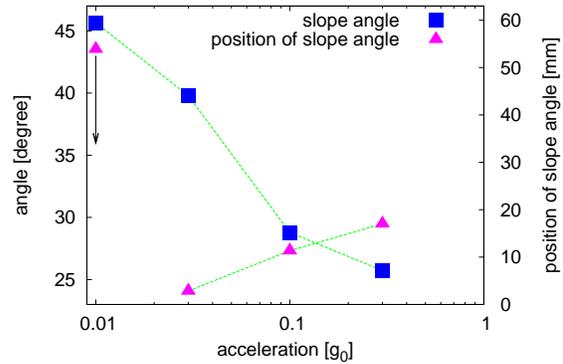}
  \caption{Position and angle of the steepest slope of the surface,
	measured in the avalanche experiments with glass beads from the top of the ridge, as a function of the gravity level. Values
	for the lowest gravity setting are measured at the end of the 4.7~s flight, the arrow indicates estimated value.}
  \label{frictionangleglas}
\end{figure}

Fig. \ref{frictionangleglas} shows that the steepest angle of the slope increases with decreasing
gravity. Due to the S-shape of the slope of the surface, increasing gravity shifts the position of the steepest part of the slope
 further away from the top of the ridge.

%%%%%%%%%%%%%%%%%%%%%%%%%%%%%%%%%%%%%%%%%%%%

\section{\label{exp2}Hourglass Experiment}

\begin{figure}
  \includegraphics[width=0.9\columnwidth]{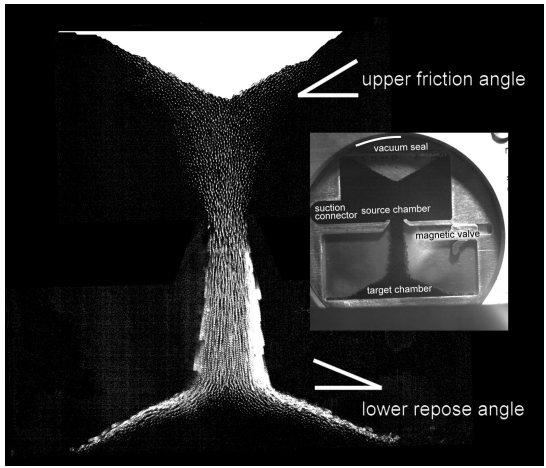}
  \caption{Software output of the hourglass experiment. Empty parts are white, filled parts
  are black and flowing particles
  appear in gray. The upper friction angle and the lower repose angle are indicated.
  The inset shows an image of the hourglass experiment, performed at a gravity level of 0.01 
  $g_0$ with basalt spheres. }
  \label{sanduhr}
\end{figure}

	The hourglass consists of two chambers separated by a magnetically
 operated valve (Fig. \ref{sanduhr}). The upper chamber is filled with the granular-matter sample.
 Upon opening the valve, the granular particles trickle into the chamber below.

	The upper chamber has a width of 50~mm.
 To ensure that the accumulated material does not reach the wall
 of the chamber too early, the lower chamber has a width of 70~mm.
 The depth ($\rm 3^{rd}$ dimension) of the hourglass is 5~mm.
 The walls are made of aluminum and glass.
	The backside features a vacuum connector to evacuate the
 experiment, which is also used to fill the upper chamber with the
 granular matter. The system could only be evacuated before the
 flight due to operational restrictions.
	The vacuum connector was connected to the outside of the drop capsule
 residing in the evacuated drop tower. For safety reasons, this connection
 had to be closed during the flight.
	During the flight, the pressure increased from 0.2~mbar to 1~mbar
	as the rotary feedthrough through
 the centrifuge was not fully vacuum-tight.

	The hourglass was mounted on the centrifuge with its top facing
 towards the center and the glass of the window in parallel to the
 rotational axis to avoid a significant influence of coriolis force.
	A force on the grains towards the smooth surface of the window is
 preferred over one perpendicular to the viewing direction.
% \cite{Brucks2008}.

	The experiment was recorded with a \emph{Voss\-k\"{u}h\-ler HCC-1000}
 C-MOS high-speed camera. The camera records 231 images per second with a
 resolution of 1024x1024 8-bit black-and-white pixels.
	Fig. \ref{sanduhr} shows an image close to the end of a flight at
 0.01 $g_0$.

\begin{figure}
  \includegraphics[height=0.9\columnwidth,angle=270]{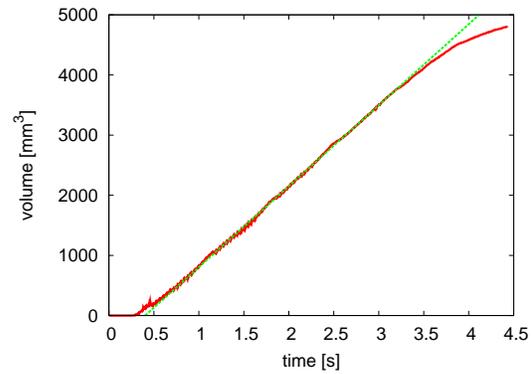}
  \caption{The measured volume drained from material in the upper chamber of the hourglass experiment with basalt spheres as a function if time, 
  measured at an acceleration level of 0.1~$g_0$. The hourglass begins to expire at 3~s.}
  \label{basaltflow01g}
\end{figure}

\begin{figure}
  \includegraphics[height=0.9\columnwidth,angle=270]{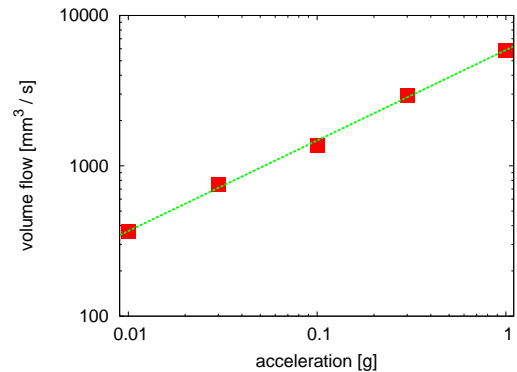}
  \caption{The volume flow of basalt spheres in the hourglass experiment as a function of the gravity level.}
  \label{basaltflow}
\end{figure}

	The camera records each pixel at a different point of time (rolling
 shutter). Due to the 50-Hz oscillation of commercial power used to power the
 lights each pixel is subject to different lighting conditions.
	A computer program was written to correct the non-constant lighting and
 to allow precise analysis of the data. A screenshot of the software is
 also shown in Fig. \ref{sanduhr}. The area drained from basalt grains is shown in white. The size of
 this area is measured to calculate the volume flow
 as shown in Figs. \ref{basaltflow01g} and \ref{basaltflow}. In areas of moving material, the brightness indicates the
velocity of the particles. Velocities exceeding 1 pixel per frame (20~mm/s)
 cannot be distinguished by the evaluation algorithm. % are indistinguishable
 The software also evaluates both (left and right)
 repose angles in both (upper and lower) chambers (see Fig. \ref{sanduhr}).

\begin{figure}
  \includegraphics[height=0.9\columnwidth,angle=270]{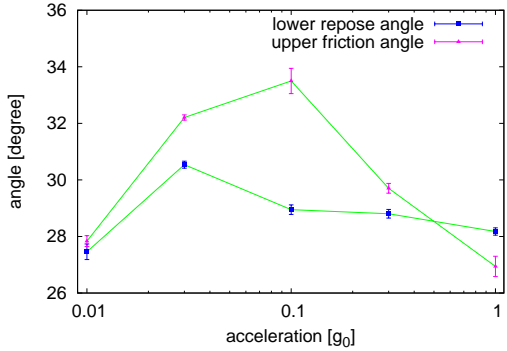}
  \caption{Repose angle in the bottom chamber and friction angle in the top chamber
  of the hourglass experiments with basalt spheres as a function of gravity level.
  Error bars denote the range between left and right angles.}
  \label{basaltangle}
\end{figure}

	The volume flow of the grains is perfectly constant over time
 and depends on the gravity level.
	According to theory, the volume flow depends on the square root of the
 gravity \cite{Hagen}. As shown in Fig.\ref{basaltflow}, our measurements do not fully agree. The exponent we found is 0.60, slightly 
 but significantly larger than theoretically predicted, and in agreement with earlier work \cite{Brucks2008}. The average values of the repose angles measured in our experiments are shown in Fig. \ref{basaltangle}. The repose angles in the
 lower chamber were determined before the accumulated material reached the walls of the chamber. The friction angle measured in the upper chamber shows a stronger dependency on the gravity level than the angle measured in the lower chamber. Surprisingly, the slope angle in the upper chamber has its maximum
 at 0.1 $g_0$. 
 
Due to the high recording rate of the camera, it was possible to track the trajectory of individual basalt particles in the hourglass experiments. From the trajectories derived by image analysis, the flow velocity within and the  total depth of the flow layer could be determined.

\begin{figure}
  \includegraphics[height=0.9\columnwidth,angle=270]{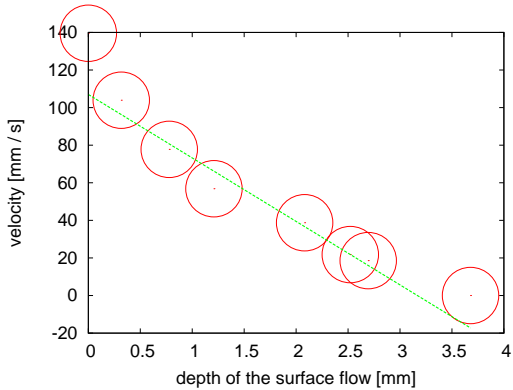}
  \caption{Surface velocity of basalt spheres in the hourglass experiment at an acceleration level of 0.3~$g_0$.
  The velocity of each layer depends on the depth of the layer. The circles
  represent the size of the grains.}
  \label{schicht03}
\end{figure}

\begin{figure}
  \includegraphics[height=0.9\columnwidth,angle=270]{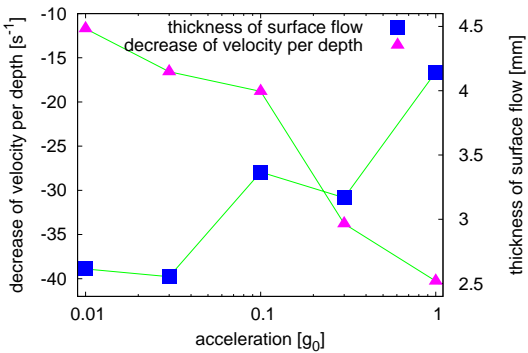}
  \caption{Decrease of the velocity per depth and the size of the
			flow in the hourglass experiments with basalt spheres as a function of the gravity level.}
  \label{flux}
\end{figure}

	Fig. \ref{schicht03} shows the speed of the grains in each
 layer at a gravity level of 0.3~$g_0$ for the basalt spheres.
	The top-most layer is excluded for the determination of the
 velocity dependence as it mainly contains ``tumbling'' and, thus, 
 not flowing grains. The non-moving particles were also excluded
 from the analysis. The resulting data (Fig. \ref{flux}) show two
 characteristic properties:
 The decrease of the flow velocity with depth is highest at high
 acceleration levels and decreases towards low gravity values.
 At the same time, the depth of the flow increases with increasing
 gravity level.  

	As both, the normal force and the gravitational force, are proportional to the gravity level, the decay of the flow velocity with depth
	is expected to be independent of gravity. Measurements, however, do clearly not support this idealized picture. The difference between theory and experiment may be caused by a friction force increasing more than linearly with the normal force. 

It must be clear that, as the depth of the surface flow strongly depends on the global
 volume flow of material through the hourglass (which
 depends on gravity), the gravity-dependence of intrinsic flow
 properties cannot easily be derived.

%%%%%%%%%%%%%%%%%%%%%%%%%%%%%%%%%%%%%%%%%%%%

\section{Summary And Outlook}

	The measurements done with the avalanche box yielded
 results predicted by common sense: Increasing gravity
 results in faster avalanches and shallower slopes for
 materials of small spherical grains.

	The volume flow through the hourglass showed a behavior unexpected
 by theory. Air pressure or humidity can be excluded as possible
 causes. Earlier experiments done under ambient pressure showed
 similar results \cite{Brucks2008}.
	The measurements' deviation from the theory for the volume flow and its dependence on gravity is still unknown.
 The exponent of 0.60 for basalt significantly exceeds the expected value
 of 0.50. The exponent is even higher when the sand sample is used.

	Size, velocities and repose angles of the surface flow
 showed clear dependencies on gravity. Theoretical expectations were
 not fully met. This may have been caused by a dependency on
 the global volume flow through the hourglass, which - in this specific
 experiment - depends on gravity and thus differs for all
 all measurements.
	To further analyze the surface flow, a setup with constant
 global volume flow could be of help.

\begin{theacknowledgments}

	This work was carried out with the help of the
	Bremen drop tower staff and was funded by the
	Deutsches Zentrum f\"{u}r Luft- und Raumfahrt (DLR).

\end{theacknowledgments}



\bibliographystyle{aipprocl} % if natbib is missing

%\bibliography{staub}

%%%%%%%%%%%%%%%%%%%%%%%%%%%%%%%%%%%%%%%%%%%
% Just a reminder that you may have to run bibtex
% All of it up to \end{document} can be removed
% if you don't like the warning.
%  \typeout{******************************************}
%  \typeout{** Please run "bibtex \jobname" to optain}
%  \typeout{** the bibliography and then re-run LaTeX}
%  \typeout{** twice to fix the references!}
%  \typeout{******************************************}

%%%%%%%%%%%%%%%%%%%%%%%%%%%%%%%%%%%%%%%%%%%
%% The following lines show an example how to produce a bibliography
%% without the help of the BibTeX program. This could be used instead
%% of the above.
%%%%%%%%%%%%%%%%%%%%%%%%%%%%%%%%%%%%%%%%%%%

\end{document}